\documentclass{ws-procs975x65}
\begin{document}

\title{How Magnetic is the Neutrino?
\footnote{Talk given at the Festschrift in honour of B. H. J. McKellar
and G. C. Joshi, November 2006.  This article is based upon the
results of Refs.~\refcite{dirac} and \refcite{majorana}.}  }

\author{N. F. Bell$^\dagger$}

\address{School of Physics,\\
The University of Melbourne,\\
Victoria, 3010, Australia\\
$^\dagger$E-mail: n.bell@physics.unimelb.edu.au}

\begin{abstract}

The existence of a neutrino magnetic moment implies contributions to
the neutrino mass via radiative corrections.  We derive
model-independent \lq\lq naturalness" upper bounds on the magnetic moments
of Dirac and Majorana neutrinos, generated by physics above the
electroweak scale.  For Dirac neutrinos, the bound is several orders
of magnitude more stringent than present experimental limits.
However, for Majorana neutrinos the magnetic moment bounds are weaker
than present experimental limits if $\mu_\nu$ is generated by new
physics at $\sim$ 1 TeV, and surpass current experimental sensitivity
only for new physics scales $>$ 10 -- 100 TeV.  The discovery of a
neutrino magnetic moment near present limits would thus signify that
neutrinos are Majorana particles.

\end{abstract}

\keywords{Neutrino, magnetic moment, neutrino mass}

\bodymatter

\section{Introduction}
In the Standard Model (minimally extended to include non-zero neutrino
mass) the neutrino magnetic moment is non-zero, but small, and is
given by~\cite{Marciano:1977wx}
\begin{equation}
\mu_\nu\approx 3\times 10^{-19}\left(\frac{m_\nu}{1{\rm eV}}\right)\mu_B, 
\label{SM}
\end{equation}
where $m_\nu$ is the neutrino mass and $\mu_B$ is the Bohr magneton.
An experimental observation of a magnetic moment larger than that
given in Eq.(\ref{SM}) would thus be a clear indication of physics
beyond the minimally extended Standard Model.  Current laboratory
limits are determined via neutrino-electron scattering at low
energies, with $\mu_\nu < 1.5 \times 10^{-10} \mu_B$~\cite{Beacom} and
$\mu_\nu < 0.7 \times 10^{-10} \mu_B$~\cite{reactor} obtained from
solar and reactor experiments, respectively.  A stronger limit can be
obtained from constraints on energy loss from stars, $\mu_\nu < 3
\times 10^{-12} \mu_B$~\cite{Raffelt}.

\begin{figure}[t]
\begin{center}
\psfig{file=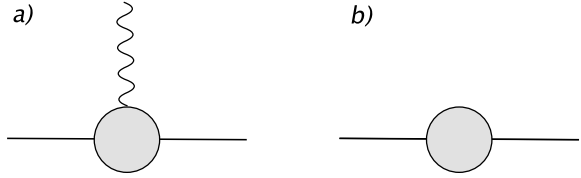,width=3in}
\end{center}
\caption{a) Generic contribution to the neutrino magnetic moment
induced by physics beyond the standard model. b) Corresponding
contribution to the neutrino mass.  The solid and wavy lines
correspond to neutrinos and photons respectively, while the shaded
circle denotes physics beyond the SM.}
\label{fig:naive}
\end{figure}

It is possible to write down a simple relationship between the size of
the neutrino mass and neutrino magnetic moment.  If a magnetic moment
is generated by physics beyond the Standard Model (SM) at an energy
scale $\Lambda$, as in Fig.~\ref{fig:naive}a, we can generically
express its value as
\begin{equation}
\mu_\nu \sim \frac{eG}{\Lambda},
\end{equation}
where $e$ is the electric charge and $G$ contains a combination of
coupling constants and loop factors.  Removing the photon from the
same diagram (Fig.~\ref{fig:naive}b) gives a contribution to the
neutrino mass of order
\begin{equation}
m_\nu \sim G \Lambda.
\end{equation}
We thus have the relationship
\begin{eqnarray}
m_\nu  \,\, \sim  \,\,  \frac{\Lambda^2}{2 m_e}  \frac{\mu_\nu}{\mu_B} 
\,\, \sim \,\,  \frac{\mu_\nu}{ 10^{-18} \mu_B}
[\Lambda({\rm TeV})]^2  \,\,\, {\rm eV},
\label{naive}
\end{eqnarray}
which implies that it is difficult to simultaneously reconcile a small
neutrino mass and a large magnetic moment.

However, it is well known that the na\"ive restriction given in
Eq.(\ref{naive}) can be overcome via a careful choice for the new
physics.  For example, we may impose a symmetry to enforce $m_\nu=0$
while allowing a non-zero value for
$\mu_\nu$~\cite{Voloshin,Georgi,Grimus,mohapatra1}, or employ a spin
suppression mechanism to keep $m_\nu$ small~\cite{Barr}.  Note though,
that these symmetries are typically broken by Standard Model
interactions.  By calculating contributions to $m_\nu$ generated by SM
radiative corrections involving the magnetic moment interaction, we
may thus obtain general, \lq\lq naturalness" upper limits on the size
of neutrino magnetic moments.

One possibility for allowing a large $\mu_\nu$ while keeping $m_\nu$
small is due to Voloshin~\cite{Voloshin}.  The original version of
this mechanism involved imposing an $SU(2)_\nu$ symmetry, under which
the left-handed neutrino and antineutrino ($\nu$ and $\nu^c$)
transform as a doublet.  The Dirac mass term transforms as a triplet
under this symmetry and is thus forbidden, while the magnetic moment
term is allowed as it transforms as a singlet.  However, the
$SU(2)_\nu$ symmetry is violated by SM gauge interactions.
For Majorana neutrinos, the Voloshin mechanism may be implemented
using flavor symmetries, such as those
in Refs.~\refcite{Grimus,Georgi,mohapatra1}.  These flavor symmetries are not
broken by SM gauge interactions but are instead violated by SM Yukawa
interactions.\footnote{We assume that the charged leptons masses are
generated via the standard mechanism through Yukawa couplings to the
SM Higgs boson.  If the charged lepton masses are generated via a
non-standard mechanism, SM Yukawa interactions do not necessarily
violate flavor symmetries.  However, such flavor symmetries must always
be broken via some mechanism in order to obtain non-degenerate masses
for the charged leptons.}

Below, we shall estimate the contribution to $m_\nu$ generated by SM
radiative corrections involving the magnetic moment term.  This allows
us to set general, \lq\lq naturalness" upper limits on the size of
neutrino magnetic moments.  For Dirac neutrinos, these limits are
several orders of magnitude stronger than present experimental
bounds~\cite{dirac}.  For Majorana neutrinos, however, the bounds are
weaker~\cite{Davidson,majorana}.

\section{Dirac Neutrinos}

We assume that the magnetic moment is generated by physics beyond the
SM at an energy scale $\Lambda$ above the electroweak scale.  In order
to be completely model independent, the new physics will be left
unspecified and we shall work exclusively with dimension $D\geq 4$
operators involving only SM fields, obtained by integrating out the
physics above the scale $\Lambda$.  We thus consider an effective
theory that is valid below the scale $\Lambda$, respects the
$SU(2)_L\times U(1)_Y$ symmetry of the SM, and contains only SM fields
charged under these gauge groups.

We start by constructing the most general operators that could give
rise to a magnetic moment operator, $\bar{\nu}_L \sigma^{\mu\nu}
F_{\mu\nu} \nu_R$.  Demanding invariance under the SM gauge group, we
have the following 6D operators
\begin{equation}
\label{eq:ops} 
{\cal O}^{(6)}_B  =  \frac{g'}{\Lambda^2}{\bar L}{\tilde \phi}
\sigma^{\mu\nu}\nu_R B_{\mu\nu}\ , \hspace{1.5cm}
{\cal O}^{(6)}_W  =  \frac{g}{\Lambda^2} {\bar L}\tau^a {\tilde \phi} 
\sigma^{\mu\nu}\nu_R W_{\mu\nu}^a\ . 
\end{equation}
where $B_{\mu\nu} = \partial_\mu B_\nu - \partial_\nu B_\mu$ and
$W_{\mu\nu}^a = \partial_\mu W_\nu^a - \partial_\nu W_\mu^a - g
\epsilon_{abc}W_\mu^b W_\nu^c$ are the U(1)$_Y$ and SU(2)$_L$ field
strength tensors, respectively, and $g'$ and $g$ are the corresponding
couplings.  The Higgs and left-handed lepton doublet fields are
denoted $\phi$ and $L$, respectively, and $\tilde\phi = i \tau_2
\phi^*$.

\begin{figure}
\label{4D}
\begin{center}
\psfig{file=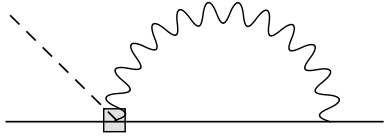,width=2in}
\end{center}
\caption{Contribution to the 4D mass operator ${\cal O}^{(4)}_M$ due to 
insertions of the magnetic moment operators ${\cal O}^{(5)}_{B,W}$.}
\end{figure}

After spontaneous symmetry breaking, both ${\cal O}^{(6)}_B$ and
${\cal O}^{(6)}_W$ contribute to the magnetic moment.  Through loop
diagrams these operators will generate contributions to the neutrino
mass.  For example, the diagram in Fig.~\ref{4D} will generate a
contribution to the neutrino mass operator, ${\cal O}^{(4)}_M={\bar
L}{\tilde \phi} \nu_R$.  Using dimensional analysis, we
estimate~\cite{dirac}
\begin{equation}
m_\nu  \,\, \sim  \,\, \frac{\alpha}{16\pi} 
\frac{\Lambda^2}{m_e}  \frac{\mu_\nu}{\mu_B}\,\, , 
\end{equation}
and thus
\begin{equation}
\mu_\nu \lesssim  3 \times 10^{-15} \mu_B  
\left(\frac{m_\nu}{1\ {\rm eV}}\right)
\left(\frac{1\ {\rm TeV}}{\Lambda}\right)^2 \,\,.
\end{equation}
If we take $\Lambda \simeq$ 1 TeV and $m_\nu \lesssim $ 0.3 eV, we obtain
the limit $\mu_\nu \lesssim 10^{-15} \mu_B$, which is several orders of
magnitude stronger than current experimental constraints.  Given the
quadratic dependence upon $\Lambda$, this constraint becomes 
extremely stringent for $\Lambda$ significantly above the electroweak scale.

\begin{figure}
\begin{center}
\psfig{file=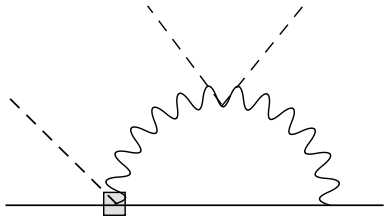,width=2in}
\end{center}
\caption{Renormalization of the mass operator, ${\cal O}^{(6)}_M$, due
to insertions of ${\cal O}^{(6)}_{B,W}$.}
\label{6Dfigure}
\end{figure}

However, if $\Lambda$ is not significantly larger that the EW scale,
higher dimension operators are important, and their contribution to
$m_\nu$ can be calculated in a model independent way.  Through
renormalization, both ${\cal O}^{(6)}_B$ and ${\cal O}^{(6)}_W$ will
generate a contribution to the 6D neutrino mass operator
\begin{equation}
{\cal O}^{(6)}_M =  \frac{1}{\Lambda^2}{\bar L}{\tilde \phi}\nu_R \left(\phi^\dag\phi\right) \ , 
\end{equation}
via the diagrams in Fig.~\ref{6Dfigure}.  Solving the renormalization group
equations we find that for $\Lambda \gtrsim 1~{\rm TeV}$,
\begin{equation}
 \label{eq:massbound}
 \mu_\nu \lesssim 8\times 10^{-15} \mu_B
 \left(\frac{m_\nu}{1\ {\rm eV}}\right) \ ,
 \end{equation}
in the absence of fine tuning~\cite{dirac}.

\section{Majorana Neutrinos}

We have seen above that the ``naturalness'' bounds on the magnetic
moments of Dirac neutrinos are significantly stronger than present
experimental limits.  However, the analogous bounds for Majorana
neutrinos are much weaker.  The case of Majorana neutrinos is more
subtle, due to the relative flavor symmetries of $m_\nu$ and $\mu_\nu$
respectively.  Majorana neutrinos cannot have diagonal magnetic
moments, but are permitted non-zero transition moments.  The
transition magnetic moment $\left[\mu_\nu\right]_{\alpha\beta}$ is
antisymmetric in the flavor indices $\{\alpha,\beta\}$, while the mass
terms $[m_\nu]_{\alpha\beta}$ are symmetric.  These different flavor
symmetries play an important role in our limits, and are the origin of
the difference between the magnetic moment constraints for Dirac and
Majorana neutrinos.

As before, we write down the most general set of operators that can
give rise to neutrino magnetic moment and mass terms, while respecting
the SM gauge group.  In the case of Majorana neutrinos, the lowest
order contribution to the neutrino mass arises from the usual five
dimensional operator containing Higgs and left-handed lepton doublet
fields:
\begin{equation}
\left[O_M^{5D}\right]_{\alpha\beta}\,=\,
\left(\overline{L^c_\alpha}\epsilon \phi\right)\left(\phi^T\epsilon
L_\beta\right),
\label{OM5}
\end{equation}
where $\epsilon = - i \tau_2$, $\overline{L^c}=L^TC$, $C$ denotes
charge conjugation, and $\alpha$, $\beta$ are flavor indices.  The
lowest order contribution to the neutrino magnetic moment arises from
the following dimension seven operators,
\begin{eqnarray}
\left[O_B\right]_{\alpha\beta}&=& g' 
\left(\overline{L^c}_\alpha\epsilon \phi\right)\sigma^{\mu\nu}
\left(\phi^T\epsilon L_\beta\right)B_{\mu\nu}, 
\label{OB}\\
\left[O_W\right]_{\alpha\beta}&=& g
\left(\overline{L^c_\alpha}\epsilon \phi\right)\sigma^{\mu\nu}
\left(\phi^T\epsilon \tau^a L_\beta\right)W_{\mu\nu}^a,
\label{OW}
\end{eqnarray}
and we also define a 7D mass operator as
\begin{equation}
\left[O_M^{7D}\right]_{\alpha\beta} = 
\left(\overline{L^c_\alpha}\epsilon \phi\right)
\left(\phi^T\epsilon L_\beta\right) \left(\phi^\dagger \phi \right).
\label{OM7}
\end{equation}

Operators $O_M^{5D}$ and $O_M^{7D}$ are flavor symmetric, while $O_B$
is antisymmetric. The operator $O_W$ is the most general 7D operator
involving $W_{\mu\nu}^a$.  However, as it is neither flavor symmetric
nor antisymmetric it is useful to express it in terms of operators
with explicit flavor symmetry, $O_W^\pm$, which we define as
\begin{eqnarray}
\left[ O_W^\pm \right]_{\alpha\beta} &=& \frac{1}{2} \left\{
\left[O_W\right]_{\alpha\beta} \pm \left[O_W\right]_{\beta\alpha} \right\}.
\end{eqnarray}

Our effective Lagrangian is therefore
\begin{eqnarray}
{\cal L} &=& \frac{C_M^{5D}}{\Lambda} O_M^{5D} 
+ \frac{C_M^{7D}}{\Lambda^3} O_M^{7D}
+  \frac{C_{B}}{\Lambda^3} O_{B} 
+\frac{ C_{W}^+}{\Lambda^3} O_{W}^+ 
+  \frac{C_{W}^-}{\Lambda^3} O_{W}^-+\cdots \ \ .
\end{eqnarray}
After spontaneous symmetry breaking, the flavor antisymmetric
operators $O_B$ and $O_W^-$ generate a contribution to the magnetic
moment interaction $\frac{1}{2} \left[\mu_\nu\right]_{\alpha\beta}\,
\overline{\nu^c}_\alpha \sigma^{\mu\nu} \nu_\beta F_{\mu\nu}$, given by 
\begin{equation}
\frac{\left[\mu_\nu\right]_{\alpha\beta}}{\mu_B} = \frac{2m_e v^2}{\Lambda^3} 
\left(\left[C_B(M_W)\right]_{\alpha\beta} 
+ \left[C_W^-(M_W)\right]_{\alpha\beta}\right),
\label{munu}
\end{equation}
where the Higgs vacuum expectation value is $\langle \phi^T \rangle =(0,
v/\sqrt{2})$.  Similarly, the operators $O_M^{5D}$ and $O_M^{7D}$
generate contributions to the Majorana neutrino mass terms,
$\frac{1}{2}\left[m_\nu\right]_{\alpha\beta}\overline{\nu^c}_\alpha
\nu_\beta$, given by
\begin{equation}
\frac{1}{2}\left[ m_\nu \right]_{\alpha\beta} 
= \frac{v^2}{2 \Lambda} \left[C_M^{5D}(M_W)\right] 
+ \frac{v^4}{4 \Lambda^3} \left[C_M^{7D}(M_W)\right].
\label{mnu}
\end{equation}

Below, we outline the radiative corrections to the neutrino mass
operators ($O_M^{5D}$ and $O_M^{7D}$) generated by the magnetic moment
operators $O_W^-$ and $O_B$.  This allows us to determine constraints
on the size of the magnetic moment in terms of the neutrino mass,
using Eqs.(\ref{munu}) and (\ref{mnu}).  Our results are summarized in
Table~\ref{summary} below, where we have defined $R_{\alpha\beta} =
m_\tau^2/| m_\alpha^2 - m_\beta^2|$, with
$m_\alpha$ being the masses of charged lepton masses.  Numerically,
one has $R_{\tau e} \simeq R_{\tau \mu} \simeq 1$ and $R_{\mu e}
\simeq 283$.

\begin{table}[htbp]
\tbl{Summary of constraints on the magnitude of the magnetic moment of
Majorana neutrinos.  The upper two lines correspond to a magnetic
moment generated by the $O_W^-$ operator, while the lower two lines
correspond to the $O_B$ operator.}
{\begin{tabular}{c|c|c}
\hline\hline
i) 1-loop, 7D & 
$\mu^W_{\alpha\beta}$ & $ \leq 1 \times 10^{-10}\mu_B
\left(\frac{ \left[m_\nu\right]_{\alpha\beta}}{1~{\rm eV}}\right)
\ln^{-1}\frac{\Lambda^2}{M_W^2} R_{\alpha\beta}$ \\
ii) 2-loop, 5D & $\mu^W_{\alpha\beta}$ & $ \leq 1 \times 10^{-9}\mu_B
\left(\frac{ \left[m_\nu\right]_{\alpha\beta}}{1~{\rm eV}}\right)
\left(\frac{1~{\rm TeV}}{\Lambda}\right)^2 
R_{\alpha\beta}$ \\  
\hline
iii) 2-loop, 7D & $\mu^B_{\alpha\beta}$ & $ \leq 1 \times 10^{-7}\mu_B
\left(\frac{ \left[m_\nu\right]_{\alpha\beta}}{1~{\rm eV}}\right)
\ln^{-1}\frac{\Lambda^2}{M_W^2}
R_{\alpha\beta}$ \\
iv) 2-loop, 5D &
$\mu^B_{\alpha\beta}$  &  $\leq 4 \times 10^{-9} \mu_B
\left(\frac{ \left[m_\nu\right]_{\alpha\beta}}{1~{\rm eV}}\right)
\left(\frac{1~{\rm TeV}}{\Lambda}\right)^2 
R_{\alpha\beta}$ \\
\hline\hline
   \end{tabular}}
\label{summary}
\end{table}


\subsection{SU(2) Gauge Boson}
\label{su2}

\subsubsection{7D mass term --- $O_W$}
As the operator $O_W^-$ is flavor antisymmetric, it must be multiplied
by another flavor antisymmetric contribution in order to produce a
flavor symmetric mass term.  This can be accomplished through
insertion of Yukawa couplings in the diagram shown in
Fig.~\ref{fig:7D}~\cite{Davidson}.  This diagram provides a
logarithmically divergent contribution to the 7D mass term, given
by~\cite{Davidson}
\begin{equation}
\label{eq:owminusone}
\left[ C_M^{7D}(M_W)  \right]_{\alpha\beta}
\simeq \frac{3 g^2}{16 \pi^2}
\frac{m_\alpha^2 - m_\beta^2}{v^2} 
\ln \frac{\Lambda^2}{M_W^2}
\left[ C_W^-(\Lambda) \right]_{\alpha\beta},
\end{equation}
where $m_\alpha$ are the charged lepton masses, and the exact
coefficient has been computed using dimensional regularization.  Using
this result, together with Eqs. (\ref{munu}) and (\ref{mnu}), leads to
bound (i) in Table~\ref{summary}.

\subsubsection{5D mass term --- $O_W$}
\begin{figure}
\begin{center}
\psfig{file=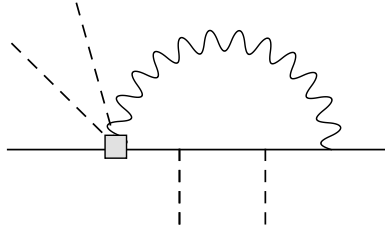,width=2in}
\end{center}
\caption{Contribution of $O_W^-$ to the 7D neutrino mass operator.}
\label{fig:7D}
\end{figure}

\begin{figure}
\begin{center}
\psfig{file=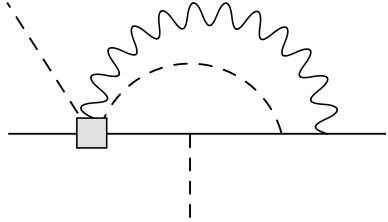,width=2in}
\end{center}
\caption{Representative contribution of $O_W^-$ to the 5D neutrino
mass operator.}
\label{fig:2loop}
\end{figure}

The neutrino magnetic moment operator $O_W^-$ will also contribute to
the 5D mass operator via two-loop diagrams, as shown in
Fig.~\ref{fig:2loop}~\cite{majorana}.  As with the diagrams in
Fig.~\ref{fig:7D}, we require two Yukawa insertions in order to obtain
a flavor symmetric result.  Using dimensional analysis, we
estimate~\cite{majorana}
\begin{equation}
\left[C_M^{5D}(\Lambda)  \right]_{\alpha\beta}
\simeq 
\frac{g^2}{(16 \pi^2)^2}
\frac{m_\alpha^2 - m_\beta^2}{v^2} 
\left[  C_W^- (\Lambda)\right]_{\alpha\beta}.
\label{5d1}
\end{equation}
This leads to bound (ii) in Table~\ref{summary}.  Compared to 1-loop
(7D) case of Eq.~(\ref{eq:owminusone}), the 2-loop (5D) mass
contribution is suppressed by a factor of $1/16\pi^2$ arising from the
additional loop, but enhanced by a factor of $\Lambda^2/v^2$ arising
from the lower operator dimension.  Thus, as we increase the new
physics scale, $\Lambda$, this two-loop constraint rapidly becomes
more restrictive.  The \lq\lq crossover" scale between the two effects
occurs at $\sim 10$ TeV.



\subsection{Hypercharge Gauge Boson}
\label{u1b}

\subsubsection{7D mass term --- $O_B$}
\begin{figure}[h]
\begin{center}
\psfig{file=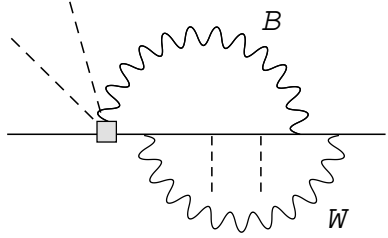,width=2in}
\end{center}
\caption{Representative contribution of $O_B$ to the 7D neutrino
mass operator at two loop order.}
\label{fig:B2M_2loop}
\end{figure}

If we insert $O_B$ in the diagram in Fig. \ref{fig:7D}, the
contribution vanishes, due to the $SU(2)$ structure of the graph.
Therefore, to obtain a non-zero contribution to $O_M^{7D}$ from $O_B$
we require the presence of some non-trivial $SU(2)$ structure.  This
can arise, for instance, from a virtual $W$ boson loop as in
Fig. \ref{fig:B2M_2loop}~\cite{Davidson}.  This mechanism gives the
leading contribution of the operator $O_B$ to the 7D mass term.  The
$O_B$ and $O_W$ contributions to the 7D mass term are thus related by
\begin{eqnarray}
\frac{(\delta m_\nu)^B}{(\delta m_\nu)^W}
\,\approx\,\frac{\alpha}{4\pi} \frac{1}{\cos^2\theta_W},
\end{eqnarray}
where $\theta_W$ is the weak mixing angle and where the factor on the
RHS is due to the additional $SU(2)_L$ boson loop.  This additional
loop suppression for the $O_B$ contribution results in a significantly
weaker neutrino magnetic moment constraint than that obtained above for
$O_W^-$.  The corresponding limit is shown as bound (iii) in
Table~\ref{summary}.

\subsubsection{5D mass term --- $O_B$}

However, the leading contribution of $O_B$ to the 5D mass term arises
from the same 2-loop diagrams (Fig.~\ref{fig:2loop}) that we discussed
in connection with the $O_W^-$ operator.  Therefore, the contribution
to the 5D mass term is the same as that for $O_W$, except for a factor
of $(g'/g)^2 = \tan^2 \theta_W$.  We thus obtain~\cite{majorana}
\begin{equation}
\left[ C_M^{5D}(\Lambda) \right]_{\alpha\beta}
\simeq 
\frac{g'^2}{(16 \pi^2)^2}
\frac{m_\alpha^2 - m_\beta^2}{v^2} 
\left[ C_B(\Lambda)  \right]_{\alpha\beta}\ \ \ ,
\label{5db}
\end{equation}
corresponding to bound (iv) in Table.~\ref{summary}.  Importantly,
this is the strongest constraint on the $O_B$ contribution to the
neutrino magnetic moment for any value of $\Lambda$, and the most
general of our bounds on $\mu_\nu^{\rm Majorana}$~\cite{majorana}.

\subsection{Comparison with experimental limits}

The best laboratory limit on $\mu_\nu$, obtained from scattering of low
energy reactor neutrinos is, $\text{``}\mu_{e}\text{''} < 0.7 \times
10^{-10} \mu_B$~\cite{reactor}. Note that this limit applies to both
$\mu_{\tau e}$ and $\mu_{\mu e}$, as the flavor of the scattered
neutrino is not detected in the experiment. Taking the neutrino mass
to be $m_\nu \lesssim 0.3$ eV (as implied by cosmological observations,
e.g. Ref.~\refcite{WMAP3yr}),  bound (iv) in Table.~\ref{summary} gives
\begin{eqnarray}
\mu_{\tau\mu},\mu_{\tau e} & \lesssim & 10^{-9} 
\left[ \Lambda(\text{TeV}) \right]^{-2}
\nonumber \\
\mu_{\mu e}  & \lesssim & 3 \times 10^{-7} 
\left[ \Lambda(\text{TeV}) \right]^{-2}.
\end{eqnarray}
For Majorana neutrinos we thus conclude that if $\mu_{\mu e}$ is
dominant over the other flavor elements, an experimental discovery
near the present limits (e.g., at $\mu \sim 10^{-11}\mu_B$) would
imply $\Lambda \lesssim 100$ TeV.  However, this would become $\Lambda
\lesssim 10$ TeV in any model in which all element of
$\mu_{\alpha\beta}$ have similar size.


\section{Conclusions}

We have discussed radiative corrections to the neutrino mass arising
from a neutrino magnetic moment coupling.  Expressing the magnetic
moment in terms of effective operators in a model independent fashion
required constructing operators containing the $SU(2)_L$ and
hypercharge gauge bosons, $O_W$ and $O_B$ respectively, rather than
working directly with the electromagnetic gauge boson.  We then
calculated $\mu_\nu$ naturalness bounds arising from the leading order
contributions to neutrino mass term, for both Dirac and Majorana
neutrinos.  
For Dirac neutrinos we found
\begin{equation}
\mu_\nu^{\rm Dirac} \lesssim  3 \times 10^{-15} \mu_B  
\left(\frac{m_\nu}{1\ {\rm eV}}\right)
\left(\frac{1\ {\rm TeV}}{\Lambda}\right)^2 \,\,,
\end{equation}
while the most general naturalness bound on the size of the Majorana
neutrino magnetic moment is
\begin{equation}
\mu_{\alpha\beta}^{\rm Majorana}\,\leq\,4 \times 10^{-9}\mu_B 
\left(\frac{\left[m_\nu\right]_{\alpha\beta}}{1~{\rm eV}}\right)
\left(\frac{1\ {\rm TeV}}{\Lambda}\right)^2 
\left| \frac{m_\tau^2}{m_\alpha^2 - m_\beta^2} \right|.
\label{generallimit}
\end{equation}
These limits can only be evaded in the presence of fine tuning.

The limit on the the magnetic moments of Dirac neutrinos is thus
considerably more stringent than for Majorana neutrinos.  This is due
to the different flavor symmetries involved, since in the Majorana
case we require the insertion of Yukawa couplings to convert a flavor
antisymmetric (magnetic moment) operator into a flavor symmetric
(mass) operator.  Our results implies that an experimental discovery
of a magnetic moment near the present limits would signify (i)
neutrinos are Majorana fermions and (ii) new lepton number violating
physics responsible for the generation of $\mu_\nu$ arises at a scale
$\Lambda$ which is well below the see-saw scale.

\section*{Acknowledgments}
This article is based upon the results of Ref.~\refcite{dirac} and
Ref.~\refcite{majorana}.  NFB thanks Vincenzo Cirigliano, Mikhail
Gorchtein, Michael Ramsey-Musolf, Petr Vogel, Peng Wang and Mark Wise
for an enjoyable and productive collaboration.

\end{document}